\documentclass[reprint,amsmath,amssymb,aps,prb,showpacs,floatfix]{revtex4-1}

\usepackage[dvips]{color,graphicx}
\DeclareGraphicsExtensions{.eps, .jpg, .png}

\usepackage{dcolumn}

\begin{document}

\title{Density functional theory study of phase IV of solid hydrogen}

\author{Chris J.\ Pickard} \email{c.pickard@ucl.ac.uk}

\author{Miguel Martinez-Canales}
\affiliation{Department of Physics \& Astronomy, University College
  London, Gower Street, London WC1E~6BT, UK}

\author{Richard J.\ Needs} 
\affiliation{Theory of Condensed Matter Group, Cavendish Laboratory,
  Cambridge CB3 0HE, UK}

\date{\today}

\begin{abstract}
  We have studied solid hydrogen up to pressures of 300 GPa and
  temperatures of 350 K using density functional theory methods and
  have found ``mixed structures'' that are more stable than those
  predicted earlier.  Mixed structures consist of alternate layers of
  strongly bonded molecules and weakly bonded graphene-like sheets.
  Quasi-harmonic vibrational calculations show that mixed structures
  are the most stable at room temperature over the pressure range
  250--295 GPa.  These structures are stabilized with respect to
  strongly-bonded molecular phases at room temperature by the presence
  of lower-frequency vibrational modes arising from the graphene-like
  sheets.  Our results for the mixed structures are consistent with
  the experimental Raman data [M.\ I.\ Eremets and I.\ A.\ Troyan,
  Nature Mater.\ \textbf{10} 927 (2011) and R.\ T.\ Howie \textit{et
    al.}\ Phys.\ Rev.\ Lett.\ \textbf{108} 125501 (2012)].  We find
  that mixed phases are reasonable structural models for phase IV of
  hydrogen.
\end{abstract}


\maketitle

\section{Introduction}

Hydrogen is the most abundant and ancient element in the Universe and
the simplest of all atoms.  
Hydrogen was 
solidified by Dewar \cite{Dewar_1899}, and Wigner and Huntington
predicted that solid hydrogen would metallize at pressures of about 25
GPa \cite{Wigner_1935}, although the metallization pressure is now
known to be much higher.  Insulating molecular phases of solid
hydrogen have been identified in static diamond anvil cell (DAC)
experiments up to pressures of over 300 GPa.  Phase I is a quantum
crystal consisting of rotating hydrogen molecules on a hexagonal close
packed lattice \cite{MaoH94}.  The orientationally ordered molecular
phases II and III \cite{Mazin_1997} appear at pressures of about 110
and 150 GPa, respectively \cite{MaoH94}, but their structures have not
been determined experimentally.  Ashcroft suggested that metallic
hydrogen might superconduct up to room temperature
\cite{Ashcroft_1968} and, more recently, possible superconducting
superfluids and metallic superfluids of hydrogen have been suggested
\cite{Babaev_2004}.  The gas giants Jupiter and Saturn and some of the
recently discovered extrasolar planets \cite{exoplanets} are believed
to contain large amounts of metallic hydrogen at high temperatures and
pressures into the terapascal range.  Metallic hydrogen has been
generated at temperatures of several thousand Kelvin in dynamic shock
wave experiments \cite{Weir_1996_shock_waves}, but it has proved very
difficult to form metallic hydrogen at lower temperatures in DACs.

DAC studies at pressures up to 300 GPa are approaching current
experimental limits.  Determining crystalline structures of hydrogen
at high pressures is extremely difficult because x-ray diffraction,
which is the workhorse of crystal structure determination, is severely
hampered by the weak scattering of hydrogen
\cite{LoubeyreLHHHMF96,Akahama_2010_x-ray}.  Raman
\cite{MaoH94,GoncharovGHM01,LoubeyreOL02} and infra-red (IR)
\cite{Hanfland_1993,MaoH94} vibrational spectroscopies have been very
important in studying the bonding and structures of hydrogen at high
pressures, although they have not proved sufficient to determine the
structures themselves.

Recent room temperature (295--300 K) DAC experiments up to pressures
of around 300 GPa have found evidence for a previously unreported
structural phase of hydrogen, that we refer to as the putative phase
IV \cite{Eremets_2011,Howie_2012}.  The intense Raman molecular vibron
mode started to drop rapidly in frequency as the pressure was
increased above about 200 GPa, and at higher pressures the samples
became dark \cite{Eremets_2011,Howie_2012}.  Darkening of samples
above 200 GPa was observed in earlier low-temperature experiments
\cite{LoubeyreOL02,Akahama_2010_raman}, suggesting a reduction in the
band gap that may be a precursor to metallization. Darkening was also
observed at high pressures in Refs.\ \onlinecite{Eremets_2011} and
\onlinecite{Howie_2012}.  Eremets and Troyan reported an increase in
electrical conductance around 260--270 GPa \cite{Eremets_2011}.  The
Raman data in Refs.\ \onlinecite{Eremets_2011} and
\onlinecite{Howie_2012} are similar, although in Ref.\
\onlinecite{Howie_2012} signals characteristic of phase IV were
obtained up to higher pressures of 315 GPa.  The Raman data of Refs.\
\onlinecite{Eremets_2011} and \onlinecite{Howie_2012} can be used to
eliminate candidate structures for phase IV.  There are differences
between these experimental data, but here we focus on quantities for
which there is a reasonable degree of consistency between the studies.

In previous work \cite{pickard_hydrogen,pickard_Brazil} we searched
for high-pressure structures of hydrogen using density functional
theory (DFT) methods and \textit{ab initio} random structure searching
(AIRSS) \cite{PickardN06_silane,Airss_review}.  Structures were found
at pressures from about 80 to nearly 400 GPa which were calculated to
be more stable than previously-known ones.  Several candidates for
phase II were found, although the enthalpy differences between them
are very small.  A single outstanding candidate of $C2/c$ symmetry was
found for phase III \cite{pickard_hydrogen}.  This structure was the
lowest in enthalpy from about 70--250 GPa when proton zero-point (ZP)
motion was included, which covers most of the region in which phase
III is observed.  The $C2/c$ structure also exhibits strong IR
activity that is characteristic of phase III
\cite{Hanfland_1993,HemleyMGM97}, and it gives a good account of the
Raman-active vibron frequency.  A molecular structure of $Cmca$
symmetry was found to be stable in the range 250--380 GPa, which we
refer to as $Cmca$-12, where the ``12'' denotes the number of atoms in
the primitive unit cell.  Above 380 GPa another structure of $Cmca$
symmetry \cite{JohnsonA00} was the most stable, which we refer to as
$Cmca$-4.  The discovery of a class of ``mixed'' hydrogen structures
consisting of alternate layers of strongly bonded molecules and
weakly-bonded graphene-like sheets was reported in Ref.\
\onlinecite{pickard_hydrogen}.  Mixed structures with $Pbcn$, $C2$ and
$Ibam$ symmetries were found and, when ZP motion was included, $Pbcn$
in particular became energetically competitive with the $C2/c$ phase
over a wide range of pressures \cite{pickard_hydrogen}.

In this work we have used AIRSS to identify candidate structures for
phase IV.  In the AIRSS approach randomly chosen structures are
relaxed to minima in the enthalpy
\cite{PickardN06_silane,Airss_review}.  The basic AIRSS method has
virtually no free parameters and is essentially unbiased, and it is
therefore an ideal starting point from which to introduce useful
biases and constraints.  The most important technical improvements on
the searches of Ref.\ \onlinecite{pickard_hydrogen} and
\onlinecite{pickard_Brazil} are that here we employ constraints to
exclude very-low-symmetry structures which, according to Pauling's
rule of parsimony \cite{Pauling_1929}, are unlikely to be low in
enthalpy, and that we have ``shaken'' low-enthalpy structures by
making small random atomic displacements and re-relaxing.  We have
also shaken supercells of low-enthalpy structures that are then able
to relax into structures with the larger supercell periodicity. AIRSS
has been applied to many systems including molecular solids such as
hydrogen \cite{pickard_hydrogen,McMahon_hydrogen_2011}, water
\cite{Pickard_water_not_h2o}, nitrogen \cite{pickard_nitrogen_2009},
ammonia \cite{Pickard_2008_ammonia} and ammonia monohydrate
\cite{Fortes_2009}.

\section{Description of the calculations}

\subsection{DFT calculations}

Our main DFT calculations were performed with the \textsc{castep}
\cite{ClarkSPHPRP05} code, the Perdew-Burke-Ernzerhof (PBE)
\cite{Perdew_1996_PBE} Generalized Gradient Approximation (GGA)
density functional, and ultrasoft pseudopotentials
\cite{Vanderbilt90}.  For the searches we used a standard ultrasoft
pseudopotential, a plane wave cut-off energy of 230 eV and a
Monkhorst-Pack \cite{Monkhorst_1976} Brillouin zone sampling grid of
spacing $2\pi \times 0.05$ \AA$^{-1}$.  The structures obtained in the
searches were re-relaxed at a higher level of accuracy consisting of a
hard ultrasoft pseudopotential, a plane wave cut-off energy of 1000 eV
and a Brillouin zone sampling grid of spacing $2\pi \times 0.03$
\AA$^{-1}$.  The enthalpies of the structures reported in the paper
were calculated at the higher level of accuracy.  For the phonon,
Raman and IR calculations we used a plane wave cut-off energy of 1200
eV and Brillouin zone sampling grids of spacing $2\pi \times 0.04$
\AA$^{-1}$.  The phonon calculations were performed with the CASTEP
code and cross-checked with the Quantum ESPRESSO code
\cite{Quantum_ESPRESSO}.

\subsection{AIRSS calculations}

\begin{figure}[ht!]
  \centering
  \includegraphics[width=0.45\textwidth]{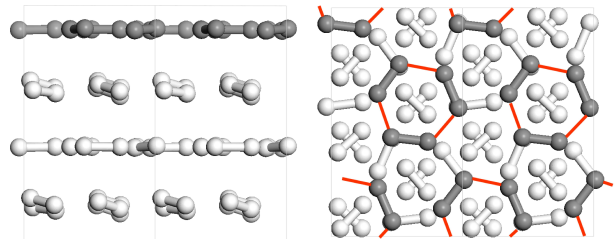}
  \caption{(color online). Side and top views of the $Pc$ structure at
    250 GPa.  The side view shows the four-layer repeat with a
    weakly-bonded graphene-like layer on top (shaded atoms).  The red
    lines show the next-nearest contacts within the layer.}
  \label{fig:H-Pc_phase}
\end{figure}

We have performed AIRSS runs in addition to those reported in our
earlier papers \cite{pickard_hydrogen,pickard_Brazil}.  The searches
were performed at 250 GPa unless otherwise specified.  Searches in 24
and 48 atom cells using symmetry constraints with from 2 up to 12
symmetry operations were performed, and 4956 relaxed structures were
obtained.  These structures included the important $Pc$ structure
(Fig.\ \ref{fig:H-Pc_phase}) which was found in searches with 2
symmetry operations.  The 16-atom $Ibam$ mixed structure
\cite{pickard_hydrogen} was shaken in supercells with up to 96 atoms,
and 305 relaxed structures were obtained.  The 48-atom $Pbcn$ mixed
structure was shaken in supercells with up to 192 atoms, giving 156
relaxed structures, and the $Pc$ structure was also found in these
searches.  The 24-atom $C2/c$ structure was shaken in supercells with
up to 96 atoms and 253 relaxed structures were obtained, and the
48-atom $Pc$ structure was shaken in supercells with up to 192 atoms
at 300 and 350 GPa, giving 333 relaxed structures.  We also found the
$Cmca$-12 and $Cmca$-4 structures in our searches.  Note that the
$C2/c$ structure that is our best candidate for phase III has 24 atoms
in the primitive unit cell \cite{pickard_hydrogen} and that it is
different from the $C2/c$ structure found by Tse \textit{et al.}\
\cite{Tse-Klug}, which has 12 atoms in the primitive unit cell.  In
the supplementary information for Ref.\ \onlinecite{pickard_hydrogen}
our 24-atom $C2/c$ structure was referred to by an alternative setting
of $B2/n$.  The two $C2/c$ structures differ in the stacking of
layers, and the 24-atom structure is lower in enthalpy.

\section{Results}

\subsection{Structure searches}

The searches confirmed our previous results \cite{pickard_hydrogen},
but we also found the $Pc$ symmetry mixed phase which has 48 atoms in
the primitive unit cell\cite{supplemental}.  The $Pc$ structure is similar to $Pbcn$, but
with a somewhat larger distortion of the graphene-like layers, see
Fig.\ \ref{fig:H-Pc_phase}.  Static lattice DFT calculations show $Pc$
to be about 2 meV per atom more stable than $Pbcn$, and at 200 GPa it
is only about 3 meV per atom higher in enthalpy than $C2/c$.  We
emphasize that the enthalpy differences between competing phases are
particularly small in hydrogen, with shifts of order 1 meV per atom
leading to significant changes in the numerical results.  Howie
\textit{et al.}\ \cite{Howie_2012} recognized that the Raman data for
phase IV might be explained by a mixed structure such as $Pbcn$, based
on the calculated data reported in Ref.\
\onlinecite{pickard_hydrogen}.

\subsection{Bond lengths of the structures} 

\begin{figure}[ht!]
  \centering
  \includegraphics[width=0.4\textwidth]{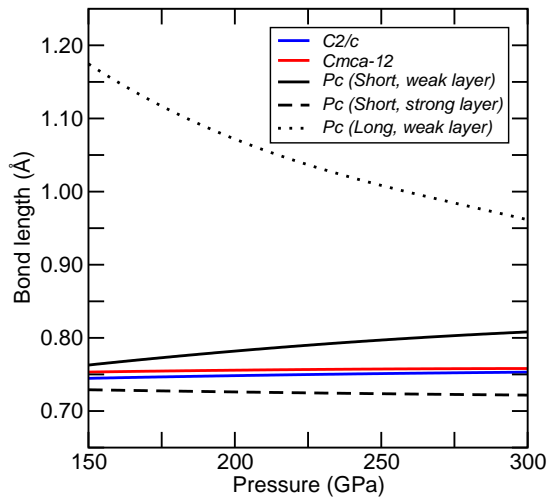}
  \caption{(color online) Evolution of the bond lengths of $C2/c$,
    $Cmca$-12 and $Pc$ with pressure. The molecular bond lengths of
    $C2/c$ and $Cmca$-12 slowly expand with increasing pressure, while
    the short bonds in the strong layers of $Pc$ undergo a slight
    contraction.  The most notable feature is the behavior of the
    bonds in the weak layers.  The short bonds in the weak layers
    expand with pressure rather more rapidly than those of $C2/c$ and
    $Cmca$-12, while the long bonds in the weak layers contract
    rapidly with pressure.}
  \label{fig:bond_lengths}
\end{figure}

As shown in Fig.\ \ref{fig:bond_lengths}, the shortest bonds in $C2/c$
and $Cmca$-12 at 150 GPa of lengths 0.745 and 0.753 \AA, respectively,
expand slowly with pressure, reaching 0.753 and 0.758 \AA\ at 300 GPa.
Fig.\ \ref{fig:H-Pc_phase} shows that the molecules in the weakly
bonded layers of $Pc$ (and similarly for $Pbcn$ and $C2$) form
distorted hexagons.  The structure evolves into the higher symmetry
$Ibam$ space group above about 350 GPa, in which the weakly bonded
layers adopt the graphene structure.  The shortest bond in $Pc$ at 150
GPa of length 0.729 \AA\ slowly contracts with increasing pressure,
reaching 0.722 \AA\ at 300 GPa.  The bond lengths in the weakly bonded
layers of $Pc$, however, change more rapidly with pressure.  At 150
GPa the weakly bonded layers have first and second nearest-neighbor
distances of 0.763 and 1.175 \AA, respectively, but upon further
compression the shorter bonds expand and the longer ones contract, and
at 300 GPa their bond lengths are 0.808 and 0.962 \AA, respectively.
These bond lengths become more nearly equal as the layer approaches
the graphene structure.

\subsection{Phonon calculations}

\begin{figure*}
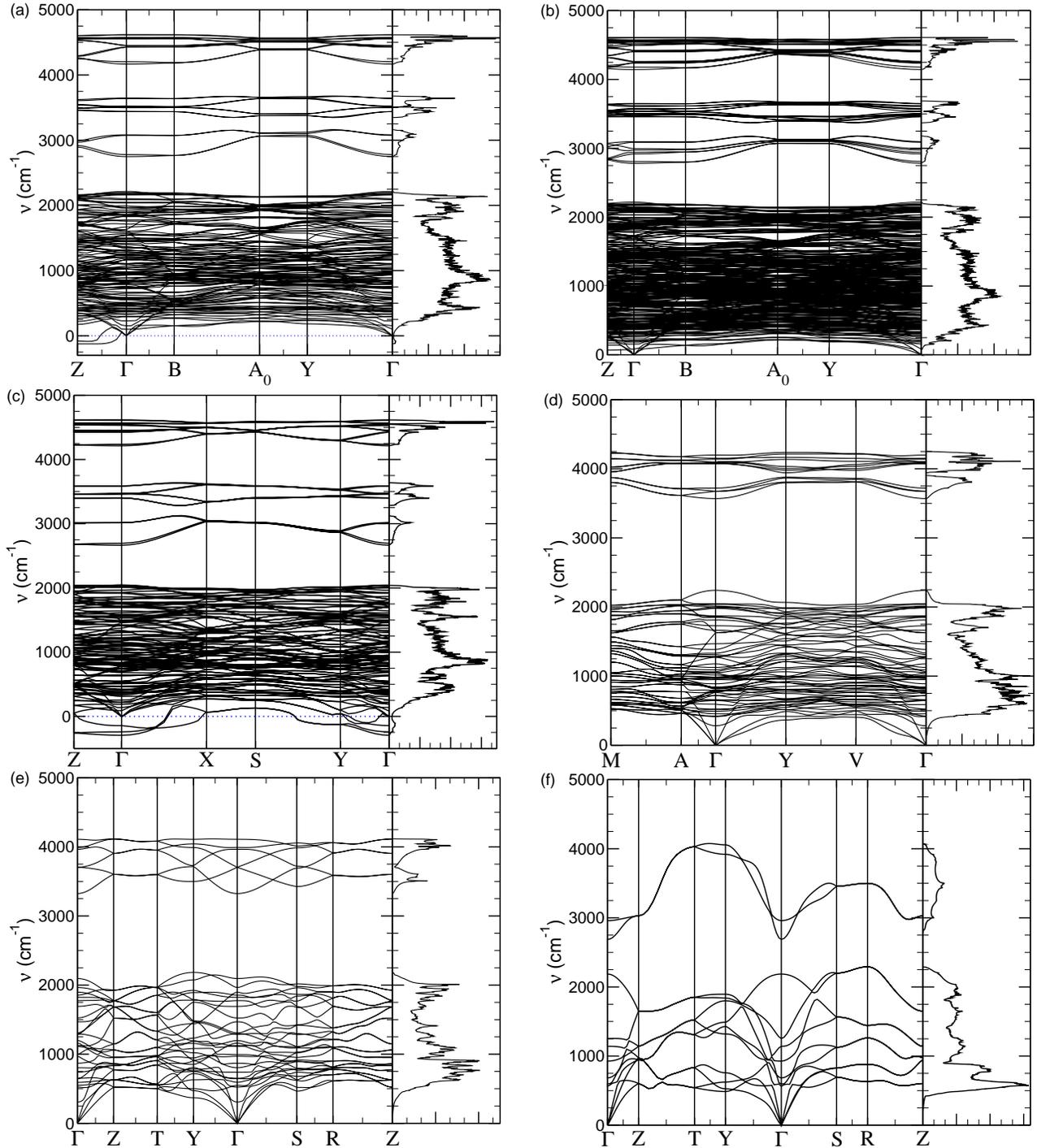

\includegraphics[width=0.45\textwidth]{Fig3a_phonons.eps}\hspace{0.5cm}
\includegraphics[width=0.45\textwidth]{Fig3b_phonons.eps}
\includegraphics[width=0.45\textwidth]{Fig3c_phonons.eps} \hspace{0.5cm}
\includegraphics[width=0.45\textwidth]{Fig3d_phonons.eps} 
\includegraphics[width=0.45\textwidth]{Fig3e_phonons.eps}\hspace{0.5cm}
\includegraphics[width=0.45\textwidth]{Fig3f_phonons.eps}
\caption{(color online) Phonon dispersion relations of (a) $Pc$, (b)
  $Pc$-96, (c) $Pbcn$, (d) $C2/c$, (e) $Cmca$-12, and (f) $Cmca$-4 at
  250 GPa.  The right hand panels show the vibrational densities of
  states.  The modes shown with negative frequencies indicate
  dynamical instabilities.  The modes below 2500 cm$^{-1}$ are lattice
  modes and those at higher frequencies are vibronic modes.  The
  densities of states of the $Pc$, $Pc$-96 and $Pbcn$ mixed phases are
  very similar and each has three distinct bands of vibrons.  $Pc$ has
  some unstable modes near Z while the unstable modes of $Pbcn$ cover
  a substantial fraction of the zone.  The $Pc$-96, $C2/c$, $Cmca$-12
  and $Cmca$-4 structures are dynamically stable. }
\label{fig:phonon_dispersion_relations}
\end{figure*}

We performed quasi-harmonic vibrational calculations for the $C2/c$,
$Cmca$-12, $Cmca$-4, $Pbcn$ and $Pc$ phases, see Fig.\
\ref{fig:phonon_dispersion_relations}.  The $C2/c$, $Cmca$-12 and
$Cmca$-4 phases are dynamically stable at 250 GPa, while the $Pbcn$
and $Pc$ phases have some unstable modes.  The unstable modes of
$Pbcn$ cover a significant portion of reciprocal space, while $Pc$ is
lower in enthalpy than $Pbcn$ and has unstable modes over only a small
region of reciprocal space.
To investigate this further we doubled the unit cell of $Pc$ to 96
atoms and re-relaxed, obtaining a $Pc$-96 structure\cite{supplemental} of the same
symmetry that is slightly lower in enthalpy than $Pc$ and has stable
phonon modes (Fig.\ \ref{fig:phonon_dispersion_relations}).  The large
size and low symmetry of the $Pc$-96 structure make it difficult to
calculate accurate vibrational spectra, and therefore we used the $Pc$
phase as our model for phase IV.

\subsection{Enthalpy-pressure relations and phase diagram}

\begin{figure}[ht!]
  \centering
\includegraphics[width=0.5\textwidth]{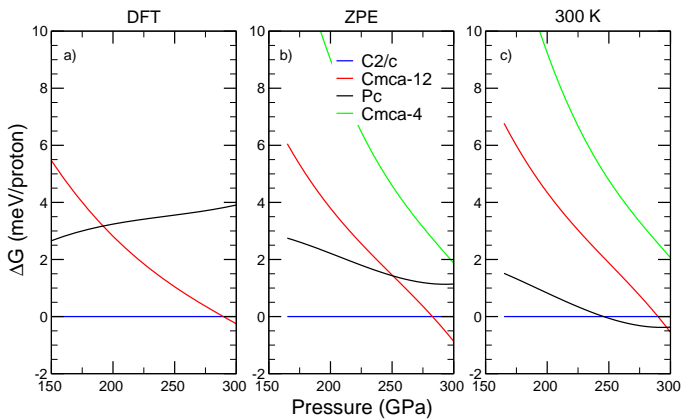}
\caption{(color online). Enthalpies/free energies of the $Cmca$-12,
  $Pc$ and $Cmca$-4 structures relative to $C2/c$ for a) the static
  lattice structures, b) with proton ZP motion and c) with full
  vibrational motion at 300 K.  The static-lattice enthalpy of
  $Cmca$-4 is too high for it to appear in a). }
  \label{fig:enthalpy-pressure}
\end{figure}

\begin{figure}[ht!]
  \centering
  \includegraphics[width=0.4\textwidth]{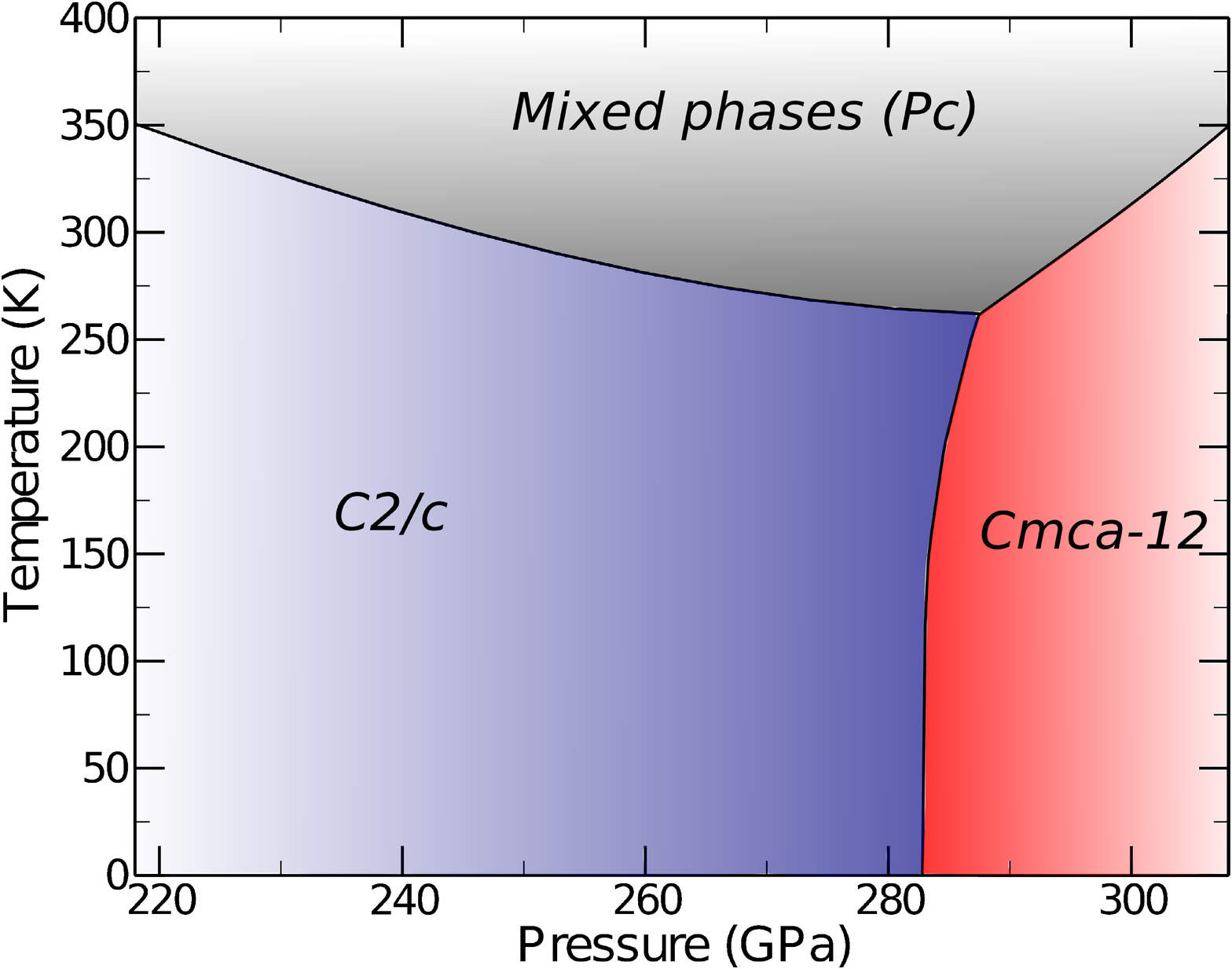}
  \caption{(color online). Computed phase diagram of hydrogen at high
    pressures.}
  \label{fig:phase_diagram}
\end{figure}

We calculated the proton ZP energy of the $C2/c$, $Cmca$-12, $Cmca$-4
and $Pc$ structures at several volumes and constructed the enthalpy at
zero temperature both with and without ZP motion, and the Gibbs free
energy at 300 K, as shown in Fig.\ \ref{fig:enthalpy-pressure}.  At
the static lattice level, $Pc$ is less stable than $C2/c$ over the
pressure range covered by (Fig.\ \ref{fig:enthalpy-pressure}a) and it
becomes less competitive compared with $C2/c$ and $Cmca$-12 as
pressure is increased.  The $Cmca$-4 phase is more than 10 meV per
proton higher in enthalpy than $C2/c$ at these pressures, and it is
therefore not visible in Fig.\ \ref{fig:enthalpy-pressure}a).
Including the vibrational ZP motion (Fig.\
\ref{fig:enthalpy-pressure}b) changes the picture significantly as the
$Pc$ and $Cmca$-4 structures become more energetically competitive
because they have lower vibrational frequencies than the strongly
molecular $C2/c$ and $Cmca$-12 phases.  Adding vibrational effects at
a temperature of 300 K (Fig.\ \ref{fig:enthalpy-pressure}c) results in
the $Pc$ phase becoming the most stable in the range 245--300 GPa.
The corresponding phase diagram (Fig.\ \ref{fig:phase_diagram}) shows
a region of stability for $Pc$ that is compatible with the
experimental data for phase IV.  Below 260 K, we also find a
transition from the $C2/c$ to $Cmca$-12 phases at about 285 GPa, which
suggests the stability of another phase of molecular hydrogen at
pressures beyond phase III.

\subsection{Raman spectra}

\begin{figure}[ht!]
  \centering
  \includegraphics[width=0.45\textwidth]{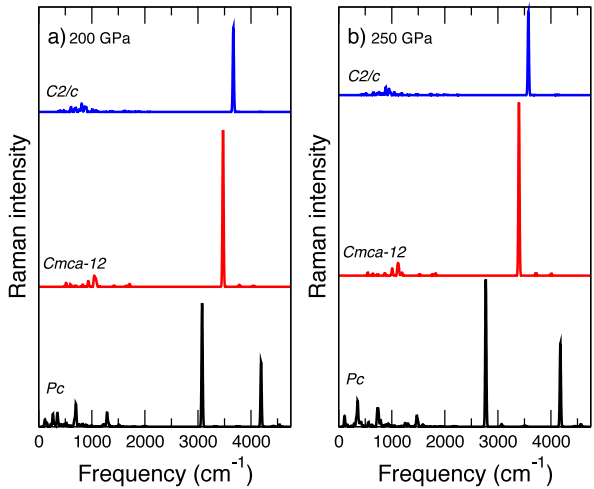}
  \caption{(color online). Raman spectra of $C2/c$, $Cmca$-12 and $Pc$
    at a) 200 GPa and b) 250 GPa.  }
  \label{fig:raman}
\end{figure}

\begin{figure}[ht!]
  \centering
  \includegraphics[width=0.35\textwidth]{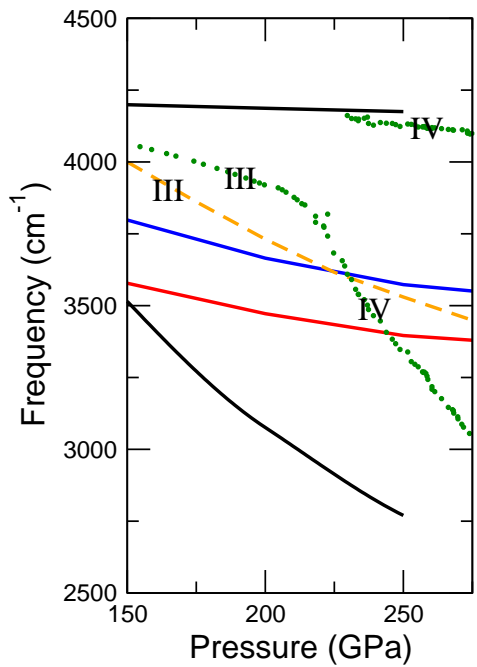}
  \caption{(color online).  Pressure dependence of the Raman vibron
    frequencies of $C2/c$ (blue solid line), $Cmca$-12 (red solid
    line) and $Pc$ (black solid line). The experimental data for phase
    III at 100 K (dashed orange line) are from Ref.\
    \onlinecite{Akahama_2010_raman} and the room temperature data for
    phases III and IV are from Ref.\ \onlinecite{Howie_2012} (dashed
    green line).}
  \label{fig:raman_vibron}
\end{figure}

To investigate the possible relationship between phase IV and the most
favorable structures obtained in our searches, we have calculated the
frequencies and intensities of the Raman modes of $C2/c$, $Cmca$-12,
$Cmca$-4 and $Pc$ \cite{Refson_2006,Porezag_1996}.  As can be seen in
Figs.\ \ref{fig:raman} and \ref{fig:raman_vibron} and Table
\ref{table:ir_data}, the $C2/c$ and $Cmca$-12 phases each have a
single intense Raman vibron mode whose frequency slowly declines with
pressure.  The pressure dependence of the Raman vibronic frequency of
$C2/c$ matches that observed for phase III reasonably well.  The
$Cmca$-4 structure \cite{JohnsonA00} is not consistent with phase IV
as it has a single intense Raman vibron mode at frequencies of about
3000 cm$^{-1}$ which slowly declines with pressure and, as mentioned
above, $Cmca$-4 is significantly less stable than other phases at 250
GPa.
$Pc$ has two strong vibronic Raman modes, one at just under 4200
cm$^{-1}$ whose frequency is very weakly pressure dependent, and a
second and more intense vibronic mode at lower frequencies which
softens rapidly with increasing pressure, see Figs.\ \ref{fig:raman}
and \ref{fig:raman_vibron} and Table \ref{table:ir_data}.  Inspection
of the vibrational eigenmodes shows that the higher frequency mode
arises from the strongly bonded molecular layers while the lower
frequency mode arises from the weakly bonded layers.  Eremets and
Troyan \cite{Eremets_2011} did not discuss the less-intense
higher-frequency Raman vibron mode of phase IV, but it is visible in
Fig.\ 2a of their paper and in Fig.\ S3 of their supplementary
information, and in the Raman data of Figs.\ 2 and 3 of Ref.\
\onlinecite{Howie_2012}.  The higher frequency vibron of the $Pc$
structure is a good match for the Raman mode of phase IV observed at
about 4130 cm$^{-1}$ in the range $\sim$220--250 GPa (see Fig.\ 2a of
Ref.\ \onlinecite{Eremets_2011}).  The higher experimental Raman
vibron frequency and that calculated for $Pc$ are almost independent
of pressure in this range, see Fig.\ \ref{fig:raman} and Table
\ref{table:ir_data}.

The experimental room temperature Raman data for hydrogen in Fig.\ 3
of Ref.\ \onlinecite{Howie_2012} shows phase IV forming from phase III
as the pressure is increased, see Fig.\ \ref{fig:raman_vibron}, which
gives rise to the ``knee'' in the lower frequency vibron branch at
around 220 GPa.  The downward slope of the room temperature
experimental data above 220 GPa \cite{Eremets_2011,Howie_2012} is in
reasonable agreement with the calculated slope of the Raman data for
the lower vibron mode of $Pc$, although the calculated modes are about
500 cm$^{-1}$ lower in frequency, see Figs.\ \ref{fig:raman} and
\ref{fig:raman_vibron} and Table \ref{table:ir_data}.  The neglect of
the effects of proton motion on the vibrational modes in our
calculations might have a significant effect on the frequencies, as
might the approximate density functional.  The $Pc$ structure does,
however, yield qualitative agreement with the two Raman vibrons
observed at room temperature in phase IV
\cite{Eremets_2011,Howie_2012}, whereas the $C2/c$, $Cmca$-12 and
$Cmca$-4 structures provide very poor descriptions.

The lattice Raman modes of $C2/c$, $Cmca$-12 and $Pc$ extend to
frequencies just above 2000 cm$^{-1}$ at 250 GPa.  $Pc$ shows
substantially higher lattice-mode Raman intensity than $C2/c$ or
$Cmca$-12.  $C2/c$ does not have any very strong lattice Raman modes,
but $Cmca$-12 has a fairly intense mode at around 1100 cm$^{-1}$.  The
strongest peak in the lattice Raman modes of $Pc$ is at 360 cm$^{-1}$
at 250 GPa, with other peaks at 110, 730, and 1470 cm$^{-1}$.  Phase
IV shows three strong lattice-mode peaks at 250 GPa \cite{Howie_2012},
although peaks below about 150 cm$^{-1}$ might be obscured by the
filter.  The strongest peak observed at 300 cm$^{-1}$ is in reasonable
agreement with our peak at 360 cm$^{-1}$, and the other two
experimentally observed modes at 630 and 1065 cm$^{-1}$ are in rough
agreement with our calculated values of 730 and 1470 cm$^{-1}$.  We
find qualitative agreement between our calculated lattice Raman modes
for $Pc$ and those observed in phase IV \cite{Howie_2012}, whereas the
$C2/c$, $Cmca$-12 and $Cmca$-4 structures provide very poor
descriptions.

\subsection{Infra-red spectra}

\begin{figure}[ht!]
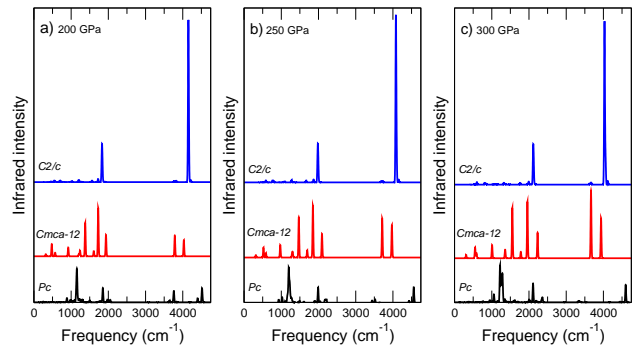

  \centering
  \includegraphics[width=0.15\textwidth]{Fig8a_ir.eps} 
  \includegraphics[width=0.15\textwidth]{Fig8b_ir.eps}
  \includegraphics[width=0.15\textwidth]{Fig8c_ir.eps}
  \caption{(color online) Infra-Red spectra of $C2/c$, $Cmca$-12 and
    $Pc$ at a) 200, b) 250, and c) 300 GPa. }
  \label{fig:infrared-spectra}
\end{figure}

We have also calculated IR spectra for $C2/c$, $Cmca$-12 and $Pc$, see
Fig.\ \ref{fig:infrared-spectra} and Table \ref{table:ir_data}.  The
vibronic modes of $C2/c$ and $Cmca$-12 soften with increasing
pressure.  $C2/c$ has a single intense IR vibron whose frequency drops
with pressure from 4150 cm$^{-1}$ at 200 GPa to 4030 cm$^{-1}$ at 300
GPa.  $Cmca$-12 has two intense IR vibrons at somewhat lower
frequencies than the single intense vibronic mode of $C2/c$.  The
frequencies of the two intense IR vibrons of $Cmca$-12 are about 3790
and 4030 cm$^{-1}$ at 200 GPa, and these frequencies fall to 3670 and
3940 cm$^{-1}$, respectively, at 300 GPa.  $Pc$ has two intense IR
vibrons at 3760 and 4520 cm$^{-1}$ at 200 GPa, but the intensity of
the lower frequency mode falls rapidly with pressure, and at 300 GPa
$Pc$ has only a single intense IR vibron peak at 4600 cm$^{-1}$.  The
higher frequency IR vibron of $Pc$ stiffens with pressure.  The
strength of the $Cmca$-12 IR vibron modes increases in comparison to
those of $C2/c$ and $Pc$ as pressure increases.  The IR vibron of
$C2/c$ is considerably more intense than those of $Pc$, and therefore
we expect a drop in IR vibron intensity at the transition from phase
III to IV.

As shown in Fig.\ \ref{fig:infrared-spectra} and Table
\ref{table:ir_data}, the IR lattice modes of $C2/c$, $Cmca$-12 and
$Pc$ stiffen with increasing pressure.  $C2/c$ has a single intense
lattice mode whose frequency of 1830 cm$^{-1}$ at 200 GPa increases to
2120 cm$^{-1}$ at 300 GPa.  This mode is in reasonable correspondence
with observations for phase III of hydrogen.  $Cmca$-12 has three
intense lattice modes at 1380, 1730 and 1940 cm$^{-1}$ at 200 GPa,
which stiffen to 1550, 1960 and 2230 cm$^{-1}$ at 300 GPa.  $Pc$ shows
two intense lattice modes at 1150 and 1850 cm$^{-1}$ which stiffen to
1250 and 2110 cm$^{-1}$ at 300 GPa.

IR data for hydrogen up to 360 GPa and 250 K has recently been
published \cite{Zha_2012}.  According to our phase diagram (Fig.\
\ref{fig:phase_diagram}) none of the data sets in the lower part of
Fig.\ 3 of Ref.\ \onlinecite{Zha_2012} are within the region of
stability of $Pc$, but the lower pressure data are within the region
of $C2/c$ and the higher pressures data at (330 GPa, 100 K) and (360
GPa, 17 K) are within the region of $Cmca$-12.  The drop in the
observed IR vibron frequency visible in Fig.\ 3 of Ref.\
\onlinecite{Zha_2012} of about 240 cm$^{-1}$ between the data at (292
GPa, 200 K) and (330 GPa, 100 K) is consistent with a phase transition
from $C2/c$ to $Cmca$-12, if the broadening of the peaks at 330 GPa
and 360 GPa is taken to arise from the unresolved pair of vibronic
peaks of $Cmca$-12, whose mean frequency is about 230 cm$^{-1}$ lower
than the vibron frequency of $C2/c$ at 300 GPa.

\begin{table*}[ht!]
  \caption{\label{table:ir_data} Frequencies in cm$^{-1}$ of the main IR 
    and Raman peaks of $C2/c$, $Cmca$-12 and $Pc$ as a function of pressure.}
\begin{ruledtabular}
\begin{tabular}{l|ccccc|ccccccc}
          &      &      &  IR  &      &      & & & & Raman & & & \\
\hline
$C2/c$    &      &      &      &      &      & & & & & & & \\
200 GPa   & 1830 & 4150 &      &      &      & 3660 & & & & & & \\ 
250 GPa   & 1980 & 4080 &      &      &      & 3570 & & & & & & \\ 
300 GPa   & 2120 & 4030 &      &      &      & & & & & & & \\ 
\hline
$Cmca$-12 &      &      &      &      &      & & & & & & & \\
200 GPa   & 1380 & 1730 & 1940 & 3790 & 4030 & 1040 & 3470 & & & & & \\ 
250 GPa   & 1470 & 1850 & 2090 & 3710 & 3970 & 1120 & 3400 & & & & & \\ 
300 GPa   & 1550 & 1960 & 2230 & 3670 & 3940 & & & & & & & \\
\hline
$Pc$      &      &      &      &      &      & & & & & & & \\
200 GPa   & 1150 & 1850 & 3760 & 4520 &      & 120 & 270 & 350 & 700 & 1290 & 3060 & 4190 \\ 
250 GPa   & 1200 & 1990 & 3520 & 4560 &      & 110 & 360 & 360 & 730 & 1470 & 2770 & 4170 \\ 
300 GPa   & 1250 & 2110 & 3340 & 4600 &      & & & & & & & \\
\end{tabular}
\end{ruledtabular}
\end{table*}

\subsection{Bandstructures}

\begin{figure}[ht!]
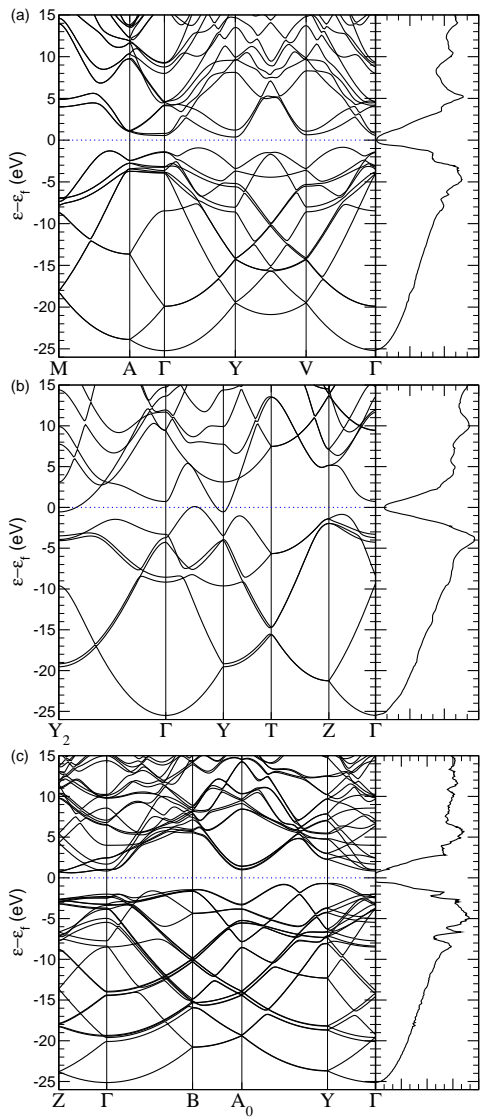

  \centering
  \includegraphics[width=0.35\textwidth]{Fig9a_bs.eps}
  \includegraphics[width=0.35\textwidth]{Fig9b_bs.eps}
  \includegraphics[width=0.35\textwidth]{Fig9c_bs.eps}
  \caption{(color online) Electronic bandstructures of $C2/c$ (top),
    $Cmca$-12 (middle) and $Pc$ (bottom) at 250 GPa.  The states below
    the blue horizontal line are occupied and those above are
    unoccupied. Note that the DFT calculations predict that $Cmca$-12
    is slightly metallic at this pressure.}
  \label{fig:bands}
\end{figure}

\begin{figure}[ht!]
  \centering
  \includegraphics[width=0.35\textwidth]{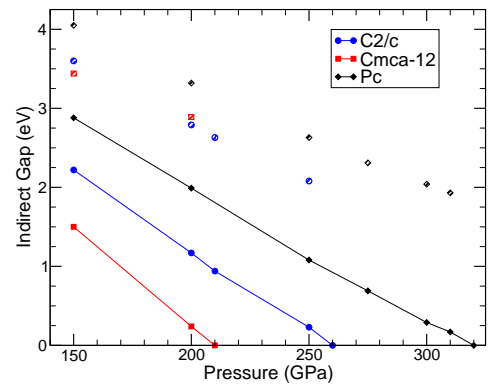}
  \caption{(color online) Band gaps of $C2/c$, $Cmca$-12, and $Pc$ as
    a function of pressure. The solid symbols show the thermal
    (minimum) band gaps while the striped symbols represent the
    minimum optical (direct in $k$-space) gaps.}
  \label{fig:band_gaps}
\end{figure}

The DFT bandstructures of $C2/c$, $Cmca$-12 and $Pc$ at 250 GPa are
shown in Fig.\ \ref{fig:bands}, and the pressure dependence of the
band gaps is shown in Fig.\ \ref{fig:band_gaps}.  $Cmca$-12 has the
smallest gap and $Pc$ has the largest.  The thermal (minimum) band
gaps decrease with pressure and go to zero at 210 GPa ($Cmca$-12), 260
GPa ($C2/c$), and 320 GPa ($Pc$), while the optical band gaps (minimum
direct gaps) are substantially larger.  Standard DFT approaches,
however, normally underestimate band gaps, and exact-exchange DFT
calculations indicate that the true gaps may be roughly 1 eV larger
\cite{StadeleM00}.  Increasing the band gaps by 1 eV gives
metallization pressures of about 250 GPa ($Cmca$-12), 310 GPa
($C2/c$), and 370 GPa ($Pc$).  We have neglected the effect of nuclear
vibrational motion on the band gaps, which is expected to be
significant.  The inclusion of vibrational motion is likely to reduce
the band gap because it amounts to averaging the gap over the
structures accessed during the motion, which are less stable than the
static lattice structure and therefore tend to have smaller band gaps.
This effect is expected to be significant in hydrogen because of the
large vibrational amplitudes, and it may be particularly important in
the weakly bonded layers of the $Pc$ phase.  The underestimation of
the band gaps arising from the PBE functional may tend to be cancelled
by our neglect of the effects of coupling between the electronic and
vibrational degrees of freedom.
We cannot therefore obtain a reliable estimate of the gap at this
level of theory but, if we suppose that the DFT calculations
underestimate the band gap by 1 eV, we find that the coupling of the
electronic and vibrational motion would have to reduce the gap by
about 1 eV to reproduce the room-temperature gap of phase IV of 1.8 eV
at 315 GPa reported by Howie \textit{et al.}\ \cite{Howie_2012}.
Theoretical work describing the anharmonic proton vibrations at finite
temperatures and their effect on the structure and band gap will be
required to investigate this in more detail.

\section{Summary and Conclusions} 

In summary, our results support the observation by Eremets and Troyan
\cite{Eremets_2011} and Howie \textit{et al.}\ \cite{Howie_2012} of a
structural phase IV of hydrogen at room temperature and pressures
above 200 GPa.  The $C2/c$, $Cmca$-12 and $Cmca$-4 candidate
structures cannot account for the observed properties of phase IV.
DFT calculations predict the $Pc$ structure to be stable at room
temperature over the pressure range 250--295 GPa, in reasonable
agreement with the experimental observations of phase IV.  Mixed
structures are stabilized by the presence of low-frequency vibrational
modes arising from the weakly-bonded graphene-like layers.  Mixed
structures explain the observation of both the high frequency Raman
vibron of phase IV that is almost independent of pressure and arises
from the short bonds in the strongly-bonded layers, and the lower
frequency vibron whose frequency falls rapidly with pressure, which
arises from the weakly bonded layers.  We find the $C2/c$ structure,
that was previously identified as a candidate for phase III
\cite{pickard_hydrogen}, to be stable at low temperatures.  The larger
lattice-mode Raman intensity of $Pc$ compared with that of $C2/c$ is
also consistent with experiment, and we find qualitative agreement
between the calculated lattice Raman modes of $Pc$ and those observed
for phase IV \cite{Eremets_2011,Howie_2012}.  We conclude that mixed
phases are reasonable structural models of phase IV of hydrogen.

\begin{acknowledgments}
  We acknowledge financial support from the Engineering and Physical
  Sciences Research Council (EPSRC) of the United Kingdom, and the use
  of the UCL Legion High Performance Computing Facility, and
  associated support services.  We thank Eugene Gregoryanz for helpful
  discussions.
\end{acknowledgments}

\end{document}